\documentclass[%
 aip, jcp,
 amsmath,amssymb,
 reprint,%
]{revtex4-1}

\usepackage{graphicx}
\usepackage{dcolumn}
\usepackage{bm}
\usepackage{longtable}
\usepackage[utf8]{inputenc}
\usepackage[T1]{fontenc}
\usepackage{mathptmx}
\usepackage{etoolbox}

\makeatletter
\def\@email#1#2{%
 \endgroup
 \patchcmd{\titleblock@produce}
  {\frontmatter@RRAPformat}
  {\frontmatter@RRAPformat{\produce@RRAP{*#1\href{mailto:#2}{#2}}}\frontmatter@RRAPformat}
  {}{}
}%
\makeatother

\begin{document}

\preprint{AIP/123-QED}

\title[]{Brute-force molecular dynamics simulation of aging in amorphous Ge$_2$Sb$_2$Te$_5$}

\author{Elisa Petroni}
\affiliation{Department of Materials Science, University of Milano-Bicocca, Via R. Cozzi 55, I-20125 Milano, Italy
}
\affiliation{STMicroelectronics, I-20864 Agrate Brianza, Italy}

\author{Dario Baratella}
\affiliation{Department of Materials Science, University of Milano-Bicocca, Via R. Cozzi 55, I-20125 Milano, Italy
}
\author{Omar Abou El Kheir} 
\affiliation{Department of Materials Science, University of Milano-Bicocca, Via R. Cozzi 55, I-20125 Milano, Italy
}
\author{Andrea Redaelli}
\affiliation{STMicroelectronics, I-20864 Agrate Brianza, Italy}
\author{Marco Bernasconi$^*$}

\affiliation{Department of Materials Science, University of Milano-Bicocca, Via R. Cozzi 55, I-20125 Milano, Italy
}

\email{marco.bernasconi@unimib.it}

\date{\today}
\begin{abstract}
Chalcogenide alloys such as the prototypical Ge$_2$Sb$_2$Te$_5$ compound are used in phase change electronic memories due to their ability to undergo a fast and reversible transition between the amorphous and crystalline phases upon heating. The amorphous phase is subject to aging that leads to an increase in the electrical resistance, which is detrimental to the operation of the devices. Using a machine learning interatomic potential, we have investigated the structural relaxations responsible for aging in amorphous  Ge$_2$Sb$_2$Te$_5$  in molecular dynamics simulations  at temperatures slightly below the glass transition. The aging simulation, lasting about 100 ns, follows an equilibration of 24 ns at 400 K. During aging we observed an enhancement of the Peierls distortion in axial bonds which is a typical feature of these alloys that we relate to the widening of the electronic band gap computed within Density Functional Theory in smaller aged amorphous models. We also observed a marked reduction of the atomic mobility with an increase of the bond lifetimes during aging. However, we do not see a reduction in homopolar bonds, which are expected to be absent in the ideal aged glass and which should be removed on a much longer time scale.

\end{abstract}

\maketitle

\section{\label{sec:intro} Introduction}
Chalcogenide alloys are used in non-volatile electronic memories (phase change memories, PCM) that rely on rapid and reversible transitions between amorphous and crystalline phases triggered by heating \cite{wuttig2007phase,noe2017phase,fantini,Cappelletti_2020,Redaelli2022, BoniardiRedaelli}. Information is encoded by exploiting the large contrast in electrical resistivity between these two phases, which can be distinguished by measuring the resistance at a low bias voltage. Higher voltage pulses are required for writing/erasing processes to either melt the crystal, which rapidly turns into the amorphous state (reset), or drive recrystallization of  the amorphous phase (set). By modulating the amorphous volume during reset or inducing partial recrystallization during set, these devices can also support multi-level or analog resistance states for neuromorphic and in-memory computing applications \cite{kuzum2012nanoelectronic,sebastianNanotech,syed}.
Because the amorphous phase is metastable, it undergoes time-dependent aging toward a more stable, defect-free glass, which causes the electrical resistivity to climb over time. This drift phenomenon follows  a power law  $R=R_o ({t}/{t_o})^{\nu}$ where $R$ and $R_o$ are the resistance at time $t$ and $t_o$ and $\nu$ is the drift exponent \cite{IelminiDrift2007}. The structural changes responsible for aging lead to a widening of the band gap, as measured by optical ellipsometry \cite{fantini_optical}, and eliminate local defects related to Urbach tails and localized in-gap states, which dictate the thermally activated sub-threshold conductivity typically modeled via the Poole-Frenkel mechanism \cite{Ielmini2009,Ielmini_Poole_Frenkel}. Gap widening and the removal of in-gap states all concur to an increase in the activation energy for conduction. Although structural relaxation is a primary explanation, alternative theories suggest that drift might stem from the thermal detrapping of carriers from the in-gap states \cite{Elliott2020, Khan2020} or alternatively from a change  with time of the Schottky barrier height for the injection of holes at the interface \cite{Yalon}, which is, however, also controlled by the evolution of the in-gap states due to structural relaxations. 

Insights into how aging affects the electronic properties of the prototypical phase-change compound Ge$_2$Sb$_2$Te$_5$ (GST)  \cite{OmarActa}  and its parent binary compound GeTe \cite{Raty2015,Gabardi2015,Zipoli2016}  (see Refs. \cite{RatyReview,MaReview} for a review),
have been gained from atomistic simulations based on density functional theory (DFT). In the DFT models of GST and GeTe  \cite{caravati2007coexistence,akola2007structural,elliot}, a small portion of Ge atoms adopt tetrahedral bonding promoted by  homopolar Ge-Ge bonds \cite{deringer2014bonding}, while the majority of atoms reside in pyramidal or defective octahedral (octahedral bonding angles, but coordination lower than six) environments.
 Using a computational alchemy technique, it was shown \cite{Raty2015}  that eliminating Ge-Ge bonds and tetrahedral Ge configurations in amorphous GeTe (a-GeTe) also  localized in-gap states are removed, while the enhancement of Peierls distortion in octahedral and pyramidal configurations leads to a widening of the band gap. Using machine learning interatomic potentials (MLIP) and metadynamics for direct aging simulations \cite{Gabardi2015}, we confirmed that removing tetrahedral Ge and homopolar bonds in a-GeTe reduces  in-gap states and  the total energy of the system.
Another study on a-GeTe  by Zipoli et al. \cite{Zipoli2016}  employed various sampling strategies to link the local structure to optical conductivity, again attributing the changes in conductivity to the reduction of the tetrahedral Ge and homopolar Ge-Ge bonds.
A correlation between drift and removal of tetrahedra was also found by Ge K-edge X-ray absorption near-edge structure spectroscopy (XANES) \cite{kolobovdrift}. Metadynamics simulations of aging in GST  \cite{OmarActa}  showed again the removal of  tetrahedra and of the corresponding in-gap states, but also other processes. Some in-gap states are removed by reducing the coordination of overcoordinated atoms, while others are removed by relaxing the environment without an apparent change in bonding of atoms on which the states are localized.

In the present work, we go a step further by simulating the structural relaxations due to aging in brute-force  MD runs that last up to 288  ns  at moderate annealing (400-450 K) below the experimental glass transition temperature (T$_g$=473 K \cite{pries2019switching}). To this end, we exploited the MLIP potential for GST devised in Ref. \cite{omar2024} and already used in our previous work on the origin of in-gap states \cite{OmarActa}. Simulations have been performed on models with 999 or 7992 atoms.  For the smaller models, we also analyzed the electronic structure using DFT calculations.

\section{Computational Details}
The MLIP for GST  was developed in our previous work \cite{omar2024}  using the neural network (NN) scheme implemented in the DeePMD package \cite{DeePMD4,DeePMD2,DeePMD3}. The NN was trained on  a DFT database of energies and forces of about 180,000 configurations of  small supercells (57-108 atoms), calculated by employing the Perdew-Burke-Ernzerhof (PBE) exchange and correlation functional \cite{PBE}.  The  potential was validated on the structural and dynamical properties of the liquid, amorphous, and crystalline phases \cite{omar2024,marcorini} and used to study the kinetics of crystallization in the bulk, in heterostructures \cite{acharya2025} and in a multimillion-atom model of a PCM cell \cite{Omar2025}. 
We added van der Waals corrections \'a la Grimme (D2) \cite{D2} that was shown to reproduce the experimental crystal growth velocity  \cite{acharya2025} and  viscosity \cite{marcorini}. 
NN-MD simulations were performed with the DeePMD code using the Lammps code as MD driver \cite{LAMMPS}, a time step of 2 fs, and  a Nos\'e-Hoover thermostat in the NVT ensemble \cite{noseart,hoover}. 
Cubic supercells of two different sizes (999 and 7992 atoms) were used at the experimental density of the amorphous phase ($\rho$=0.0309 atom/{\AA}$^3$)\cite{ghezzi}. The large amorphous GST (a-GST) models were generated by quenching from the melt which was prepared in turn from a random model initially heated at 2000 K for 20 ps and then quenched in 40 ps to 1000 K at which it was equilibrated for 20 ps. The system was then quenched to 400 K in 100 ps.  To evaluate aging, in the small model we continued the simulation at 400 K for other 288 ns.  Instead, in the large model the system was first equilibrated at 400 K for 24 ns and then the temperature was brought instantaneously to 450 K, still below the glass transition temperature T$_g$ whose last experimental estimate is 473 K \cite{pries2019switching}, to speed up the aging process and observe structural relaxation on a shorter 118 ns run. The equilibration at 400 K for 24 ns was aimed at mimicking the generation of the amorphous phase in the reset process in the memory device. The resulting amorphous phase is presumably still less equilibrated than in the device in which the reset pulse lasts about 100 ns \cite{Narayanan}.
The simulations were performed at constant volume to mimic the conditions of the real device in which the active material in confined by dielectrics and electrodes.

We checked that crystalline nuclei do not form during aging using 
the Steinhardt order parameter Q$^{dot}_{4}$, \cite{q4,q4bis} also used in our previous work \cite{omar2024}, which we refer to to for further details. We chose the threshold $Q^{dot}_{4} > 0.8$ to identify an atom as crystalline as we did in our previous work \cite{omar2024} (the correct value for the threshold used in Ref. \cite{omar2024} is also 0.8, and not 0.87 as reported by mistake in the supplementary material).

DFT-MD simulations were also performed for the smaller model using the CP2K suite of programs \cite{quickstep1,quickstep2} and the same theoretical framework used to generate the MLIP, i.e. the PBE functional and norm-conserving  pseudopotentials \cite{GTH2,GTH1}  with four, five and six valence electrons  for Ge, Sb and Te. The Kohn-Sham (KS) orbitals were expanded in a Triple-Zeta-Valence plus Polarization Gaussian-type basis set, while the charge density has been expanded in a plane-wave basis set with a cut-off of 100~Ry to efficiently solve the Poisson equation within periodic boundary conditions using the Quickstep
scheme \cite{quickstep1,quickstep2}. Brillouin Zone integration was restricted to the supercell $\Gamma$ point.
The amorphous models before and after aging were equilibrated at 300 K for 20 ps in DFT-PBE simulations. Snapshots extracted from these trajectories were then optimized at 0 K 
 with the  hybrid functional HSE \cite{hse06},  to better reproduce the band gap.
The electronic density of states (DOS) of these latter configurations was computed at the HSE level from
KS orbitals at the supercell $\Gamma$-point.

To quantify the localization properties of individual KS states, we computed the inverse participation ratio (IPR)
which is defined for the $i$th state by $\sum_j c_{ij}^4 / \bigl(\sum_j c_{ij}^2\bigr)^2$ where $j$ runs over the Gaussian-type orbitals (GTOs) of the
basis set, while $c_{ij}$ are the expansion coefficients of the $i$-th
KS state in GTOs. According to this definition, the value of the $IPR$
varies from $1/N$ for a completely delocalized KS state where $N$ is the total number of the Gaussian-type orbitals of the basis set, to one for
a mode completely localized on a single orbital.

\section{Results}

Before discussing the dynamical and structural changes due to aging, we briefly recall the structural features of a-GST as emerged from previous DFT work \cite{caravati2007coexistence,akola2007structural,elliot} and confirm by  our previous  MLIP simulations \cite{omar2024}.
Te atoms are mostly 3-coordinated in a pyramidal geometry (three bonding angles of 90$^{\circ}$), Sb atoms are both 3-coordinated in a pyramidal geometry  and 4- or 5-coordinated in a defective octahedral environment (octahedral bonding angles but coordination lower than six),  Ge atoms are mostly in pyramidal or defective octahedral geometry with a minority fraction in tetrahedral geometries favored by Ge-Ge and Ge-Sb bonds \cite{caravati2007coexistence}. A quantitative measure of the fraction of tetrahedral environments  was obtained in Ref. \cite{omar2024} from the local order parameter $q$ introduced in Ref.~\cite{qparam}. It is defined as $q=1-\frac{3}{8}\sum_{\rm i > k} (\frac{1}{3} + \cos \theta_{ \rm ijk})^2$,   where the sum runs over the pairs of atoms bonded to a central atom $j$ and forming a bonding angle $\theta_{\rm ijk}$. The order parameter is $q$=1 for the ideal tetrahedral geometry of 4-coordinated atoms, to $q$=0 for the 6-coordinated octahedral site, to $q$=5/8 for a 4-coordinated defective octahedral site, and $q$=7/8 for a pyramidal geometry. A fraction of Ge atoms in a tetrahedral geometry  of about 28\% was obtained in Ref. \cite{omar2024} by integrating the $q$-parameter for 4-coordinated Ge atoms between 0.8 and 1 as discussed in  previous works \cite{spreafico}. 
 
 Defective octahedral geometries involve axial bonds, i.e. a pair of bonds with bonding angle close to 180$^o$.  The pairs of axial bonds feature a variable degree of Peierls distortion consisting of the formation of a long and a short bond.  
DFT-MD simulations 
provided a correlation between the 
enhancement of the Peierls distortion by decreasing the temperature  and either the deepening of the pseudogap at the Fermi level in the DOS  of liquid GST above $T_m$ \cite{mazzarelloliq}, or  the metal-semiconductor transition below $T_m$ \cite{cobelli2021}. As already mentioned, the enhancement of the Peierls distortion
was correlated with the widening of the band gap in DFT calculations on a-GeTe \cite{Raty2015}.
In Ref. \cite{marcorini}, a correlation was also found between different degrees of Peierls distortion and the onset of dynamical heterogeneity (DH) in supercooled liquid GST.
The large majority of bonds in DFT models of a-GST are Ge-Te and Sb-Te bonds with a small fraction of wrong bonds (homopolar Ge-Ge, Sb-Sb, Te-Te bonds and Ge-Sb bonds \cite{omar2024}). The fraction of different bonds, the pair correlation functions and  the distribution of the coordination numbers for the different species  in the 7992-atom  model used here are reported in Ref. \cite{OmarActa} to which we refer for further details (See also Sec. 3.2). 
\begin{figure}
    \centering
    \includegraphics[width=0.45\linewidth,keepaspectratio]{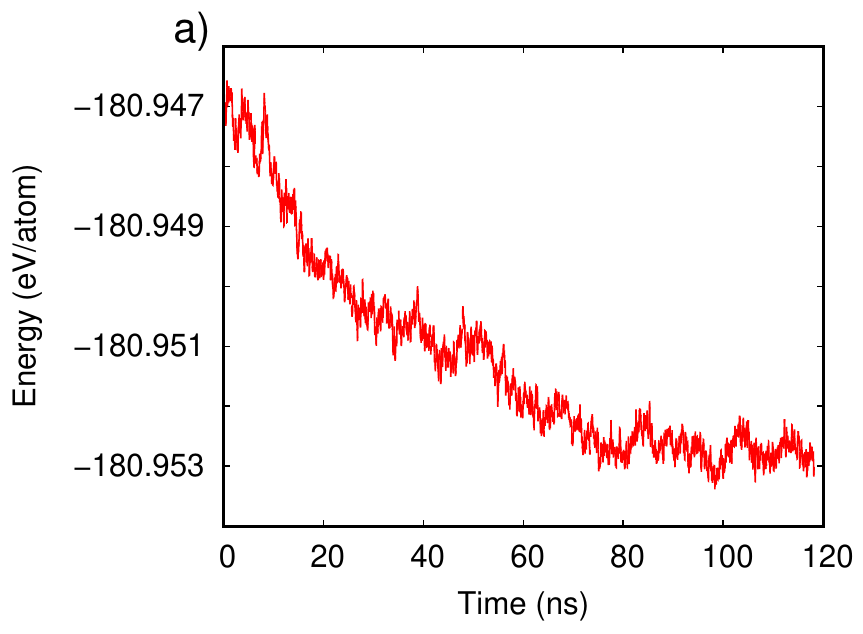}    \includegraphics[width=0.42\linewidth,keepaspectratio]{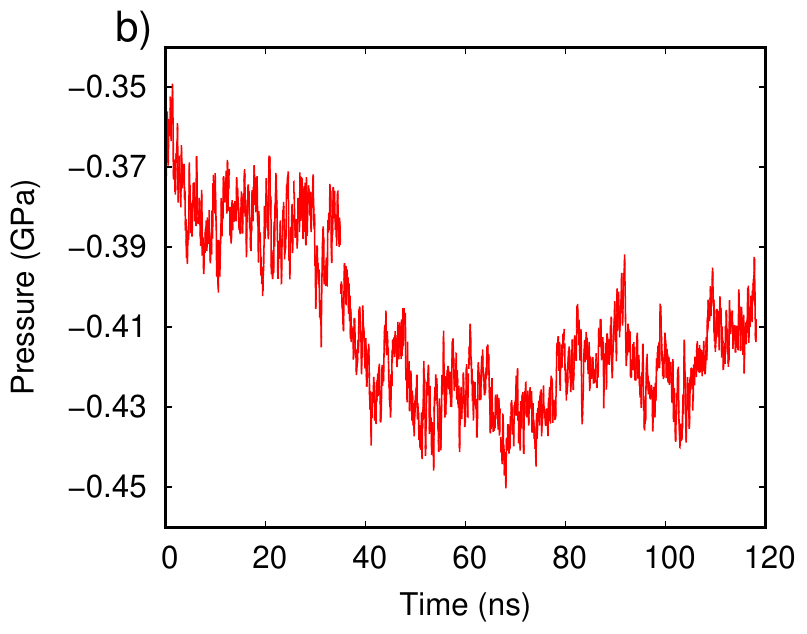}
    \caption{a) Energy per atom as a function of time in the large model at 450 K.
    b) Pressure as a function of time.}
    \label{energyA}
\end{figure}

In the following, we will discuss in separate sections the effects of aging on the dynamical, structural and electronic properties of our models of a-GST.
The results on the dynamical and structural properties (Secs. \ref{dynamical} and \ref{structure}) refer to the large model (7992 atoms at 450 K), while the results on the electronic properties (Sec. \ref{electronic}) refer to the small model (999 atoms  at 400 K).

Here, we first show the evolution in time of the energy and pressure  shown in Fig. \ref{energyA} for the large model.
The pressure is very small and negative. Since the volume is fixed at the experimental value, this  means that the equilibrium density of our MLIP-D2 is slightly larger than the experimental one. There is a slight further decrease of pressure in the first 40 ns, and then the pressure oscillates around -0.42 GPa. In contrast, the energy is continuously decreasing, with some oscillations, down to 118 ns, with a pace decreasing with time.

\subsection{Changes in dynamical properties upon aging}\label{dynamical}
\begin{figure}
    \centering
    \includegraphics[width=0.53\linewidth,keepaspectratio]{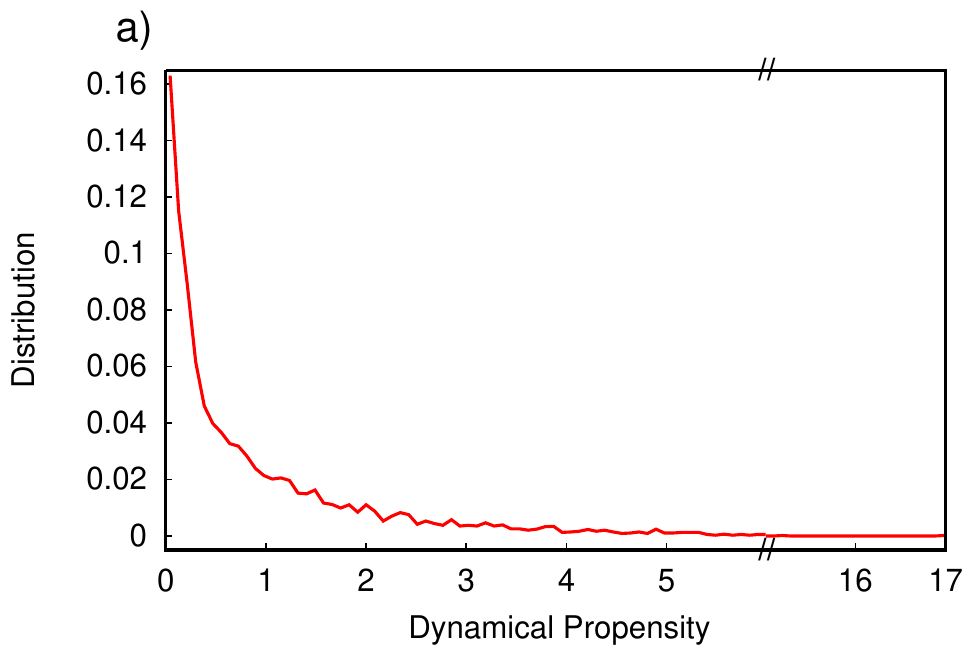}    \includegraphics[width=0.4\linewidth,keepaspectratio]{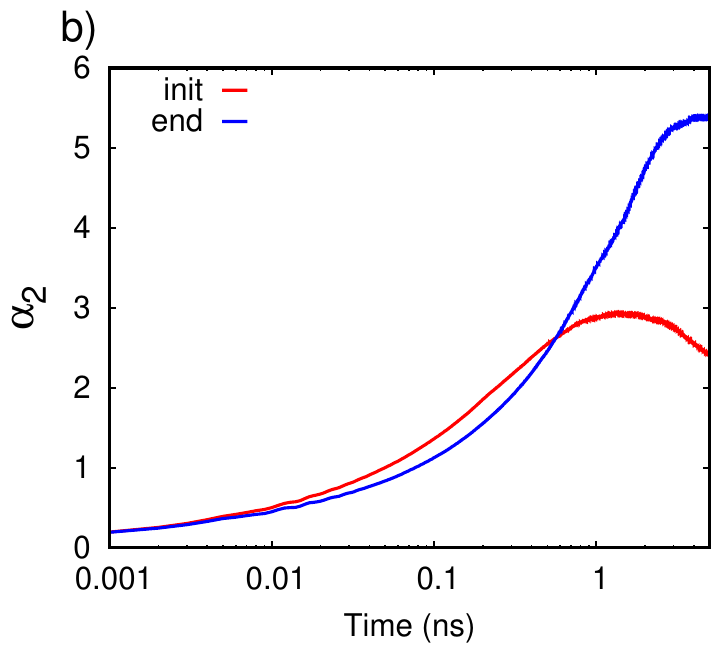}
    \caption{a) Distribution of dynamical propensity (DP, Eq. \ref{eq:dp}) for  the large model at 450 K. $\sigma$ = 1.43 and the bin size is 0.085. b) The non-Gaussianity parameter $\alpha_2$ (Eq. \ref{eq:alpha2}) at the beginning  and at the end  of the simulation of the large model at 450 K.} 
    \label{DP}
\end{figure}

Regarding changes in the dynamical properties upon aging, we first mention that a subset of atoms move considerably  and that atoms do not simply vibrate around their equilibrium position despite the relatively low temperatures (450 K) below T$_g$ at which simulations were performed. The dynamics is highly heterogeneous, with atoms moving substantially more than the average.
 We quantify the tendency of each atom $i$ to move by computing
the so called dynamical propensity (DP) over  time $t^{*}$ defined by

\begin{equation}
DP_i(t^*)=\left \langle \frac{(\vec{r}_i(t^*+t_0)-\vec{r}_i(t_0))^2}{MSD_{\alpha_i}} \right
\rangle,
\label{eq:dp}
\end{equation}

where  $\vec{r}_i$ is the position of i-th atom, $\alpha_i$ denotes the atomic species of the atom $i$ (Ge, Sb, or Te), $MSD_{\alpha_i}$ is the mean square displacement (MSD) of that species for the same time $t^*$, and $\langle\dots\rangle$ represents the average over initial times $t_0$.  The normalization factor $MSD_{\alpha_i}$ is introduced
because different species have different mobilities: atoms whose tendency to move is close to the average of their
species have $DP\approx 1$, whereas deviations from unity correspond to atoms more or less mobile than the average. 
The distribution of the DPs is shown in Fig. \ref{DP}a (see Ref. \cite{marcorini} for the corresponding plot in the liquid above T$_g$). 
DPs are computed with $t^*=$ 118 ns, i.e. the end of the simulation, and then a  single initial time 0 is used, corresponding to the beginning of the simulation. We obtain $MSD_{Ge}$=30.6 \AA$^2$, $MSD_{Sb}$=19.1 \AA$^2$, and $MSD_{Te}$= 13.6 \AA$^2$.
The dynamics is clearly heterogeneous, with a fraction of atoms moving substantially more than  the average. 
This atomic motion leads to the breaking and reforming of bonds whose lifetime will be quantified later on.
As already mentioned, DH in the supercooled liquid phase down to 550 K was also studied in our previous work \cite{marcorini}. Here, we consider the extreme case of a subset of atoms undergoing structural relaxation below T$_g$. 
The number of most mobile (MM) atoms with DP larger than 1+1.6$\sigma$, where $\sigma$ is the variance of the distribution and equal to 1.43, is about 500. From the $MSD_{\alpha_i}$ given above, this means that  Ge atoms in this class, for example, move more than 10.1 \AA.
We will consider this subset of atoms in the analysis of the structural changes induced by aging in the next section.

A common measure of  DH is the non-Gaussianity parameter $\alpha_2$
\cite{Raman}:

\begin{equation}
\alpha_2(t)=  \frac{ 3\langle(\vec{r}(t-t_0)-\vec{r}(t_0))^4\rangle}{5\langle(\vec{r}(t-t_0)-\vec{r}(t_0))^2\rangle^2}  -1
,
\label{eq:alpha2}
\end{equation}

where the average runs over different particles and different initial times $t_0$.
$\alpha_2$ is zero at $t$ = 0, then increases with time, reaching a maximum before decreasing and going to zero for $t \rightarrow \infty$. The value of the maximum is a measure
of the non-Gaussianity of the dynamics. In Fig. \ref{DP}b we show $\alpha_2$ on the time scale of a few ns computed at the beginning and at the end of simulation, which clearly indicates that DH increases during aging, as reported in the literature for several glasses (see for example Ref. \cite{schober} for Te).

\begin{figure}
    \centering  \includegraphics[width=0.7\linewidth,keepaspectratio]{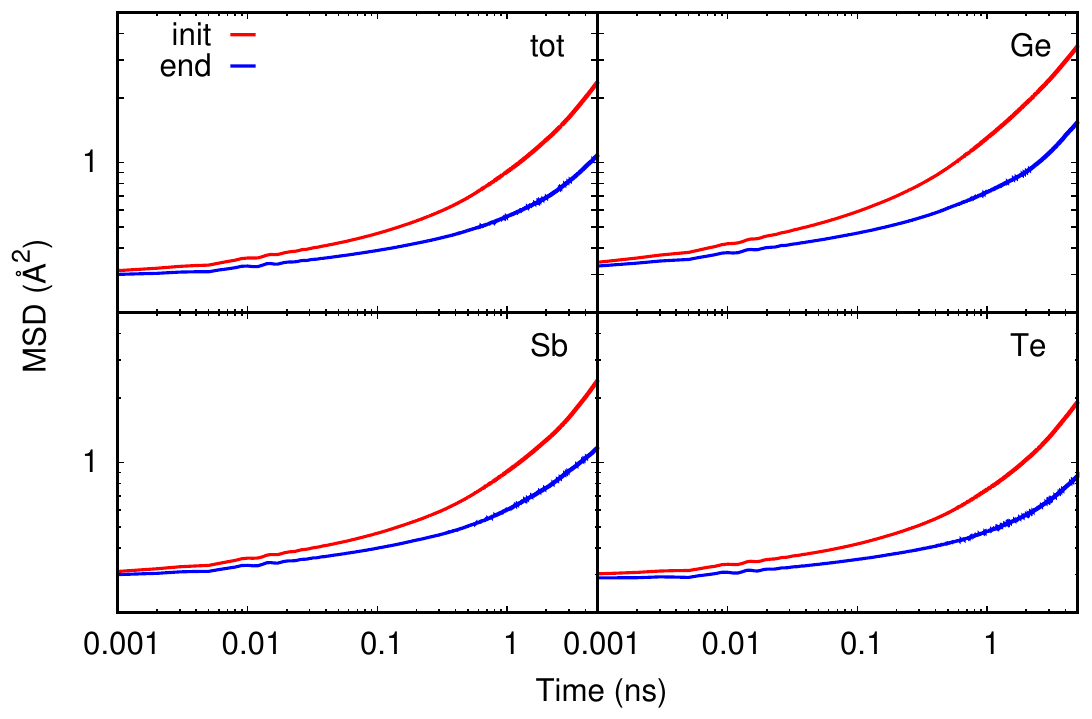}
    \caption{MSD for the different species at the beginning  and at the end  of the simulation of the large model at 450 K.} 
    \label{MSD}
\end{figure}

The increase in DH during aging is also accompanied by a slowdown of the dynamics that can be quantified, for example, by the MSD computed at the beginning and at the end of the simulation  and resolved for the different species, as  shown in Fig. \ref{MSD}. The MSD is small because it is averaged over all atoms, while, as mentioned above, only a subset of atoms move substantially and exit from the cage formed by their neighbors. Still, the reduction in MSD after aging is clear.

\begin{figure}
    \centering
    \includegraphics[width=0.42\linewidth,keepaspectratio]{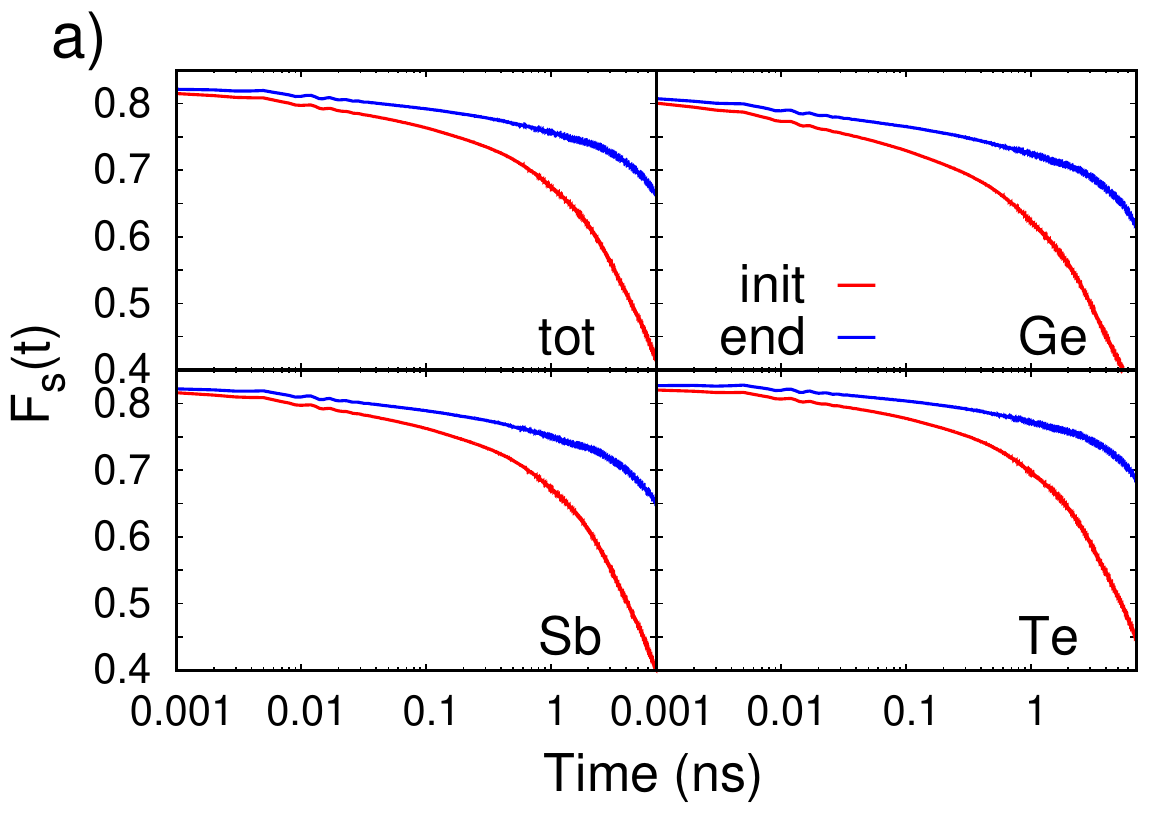}    \includegraphics[width=0.45\linewidth,keepaspectratio]{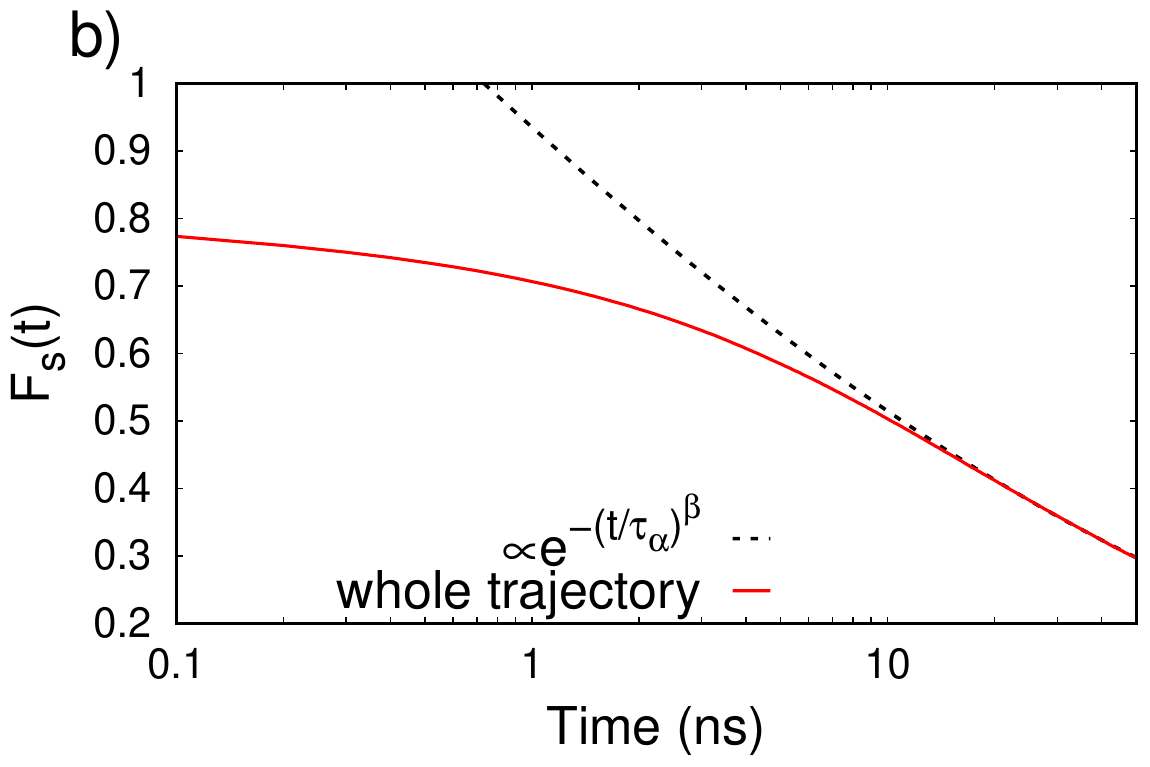}
    \caption{a) Incoherent intermediate scattering function $F_s(q_o,t)$ as a function of time, resolved for the different species, at the beginning and at the end of the simulation of the large model  at 450 K.
    b) ISF computed over the whole trajectory and the KWW fit.} 
    \label{fig:ISF}
\end{figure}

Another indication of dynamical slowdown is given by the 
(self) intermediate scattering function (ISF) $F_s(q,t)$ defined by

\begin{equation}
F_s(q,t) = \langle\Phi_s(q,t)\rangle \; ; \;
\Phi_s(q,t) = \frac{1}{N}\sum_{j}^N\exp\{i\vec{q}\cdot[\vec{r}_j(t_o) - \vec{r}_j(t+t_o)]\}
\label{eq:ISF}
\end{equation}

where the sum runs over the $N$ atoms at positions $\vec{r}_j$ at different times. The average $\langle … \rangle$ is over the initial times $t_0$. Since in a homogeneous system $F_s(q,t)$ depends only on the modulus of $\vec{q}$, we also averaged over 6 directions of  $\vec{q}$. We computed $F_s(q,t)$  at $q$ = $q_o$ = 2 {\AA$^{-1}$  } which corresponds to the position of the main peak of the static structure factor measured experimentally in Ref. \cite{mazzarelloliq}. The ISF computed on the first and last 10 ns  are shown in Fig. \ref{fig:ISF}a. It is clear that the plateau of the ISF, typical of the supercooled liquid close to T$_g$ \cite{cavagna}, extends to longer times upon aging as reported by MD simulations in many glassy systems (see for example Ref. \cite{kobepj2000} for Lennard-Jones systems \cite{kobepj2000}, Ref. \cite{schober} for Te and Ref. \cite{micoulaut2016} for a review on network glasses).
On the other hand, the ISF averaged over the entire trajectory, 118 ns long, can be fitted at long times by 
the Kohlrausch-Williams-Watts (KWW) \cite{KWW1,KWW2} stretched exponential function   $F_s(q_o,t) \sim e^{-(\frac{t}{\tau_{\alpha}})^{\beta}}$ (see Fig. \ref{fig:ISF}b).  The KWW fitting yields $\beta \simeq 0.14$ indicating that relaxations occur over different time scales (the presence of a single characteristic relaxation time would imply $\beta$=1). The low value of $\beta$ is also related to the fact that during aging the characteristic relaxation times increase. The extracted value of $\beta$ is remarkably similar to the value of 0.2 obtained in Ref. \cite{FantiniBeta} from the inclusion of  KWW relaxations in a phenomenological model of the drift with time of the threshold voltage in the selector devices of the 3DXPoint$^{\rm TM}$ devices \cite{fazio}.
Although the chalcogenides used in these devices are different from GST, this result indicates a similarity in the relaxation processes of the two classes of  materials.

\begin{figure}
    \centering
    \includegraphics[width=0.55\linewidth,keepaspectratio]{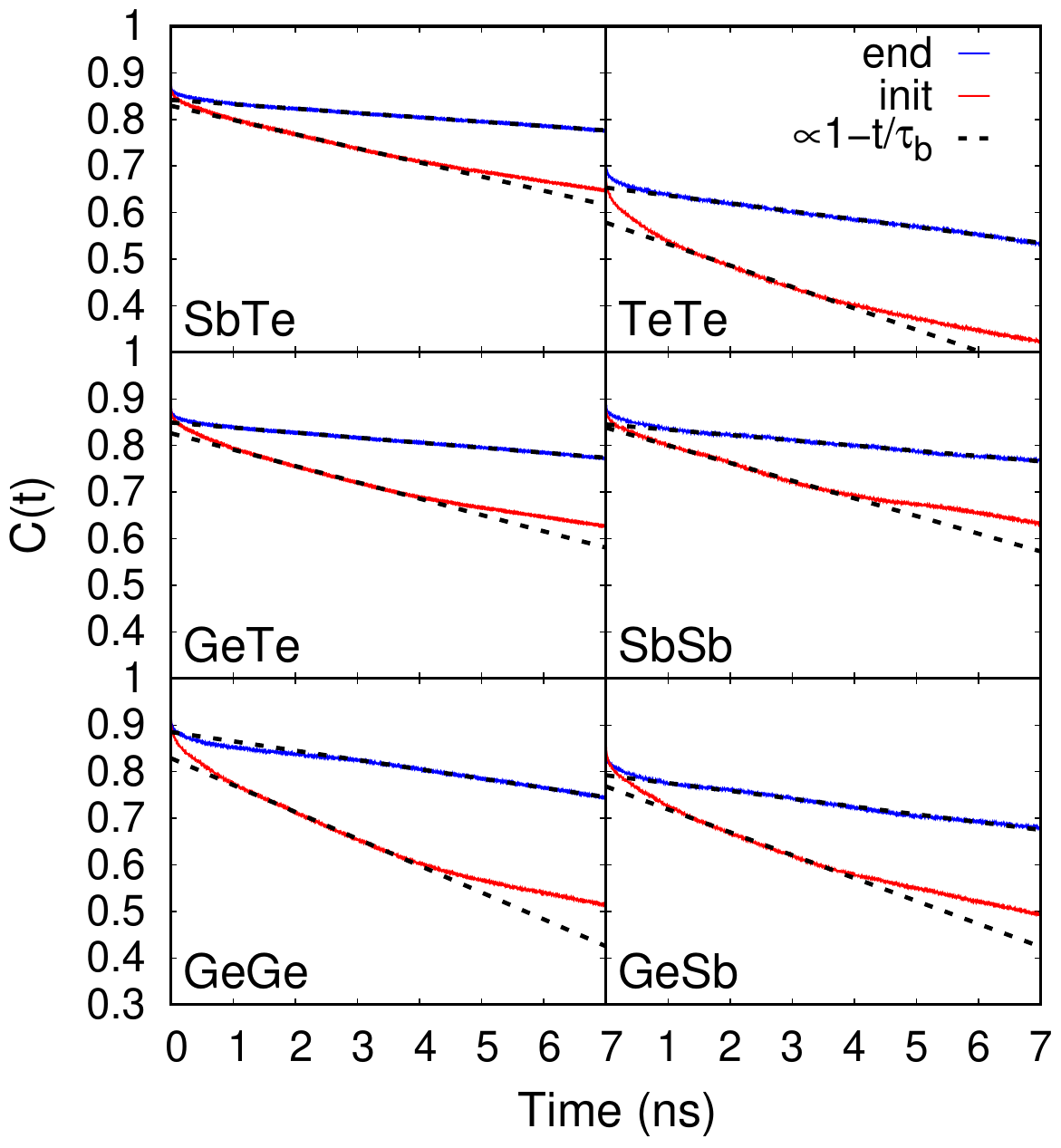}
    \caption{Bond-bond correlation function for the pair species at the beginning and the end of the simulation of the large model (118 ns at 450 K). The linear fit (Eq. \protect\ref{eq:lifetimeshort}) is shown by dashed lines.} 
    \label{C_t}
\end{figure}

\begin{table}[]
    \centering
    \begin{tabular}{c|c c c c c c}
    \hline
        & GeGe & GeSb & GeTe & SbSb & SbTe & TeTe\\
    \hline
          init & 14.4 & 15.6 & 23.6 & 22.1 & 27.3 & 12.6\\
        end & 44.3 & 47.3 & 78.9 & 74.8 & 90.2 & 38.6\\

    \hline
    \end{tabular}
    \caption{Bond lifetimes (ns) for the different pair species at the beginning and at the end of the simulation of the large model at 450 K.}
    \label{table:lifetime}
\end{table}
The slowdown in dynamics with aging can be quantified by computing the lifetime of the bonds.
Following a previous work on the study of bond lifetimes in liquid GST \cite{lee2025wrongbonds},  we studied the slow bond dynamics using the intermittent lifetime obtained from the bond-bond correlation function

\begin{equation}
C(t)= \frac{\langle h(t+t_0)h(t_0) \rangle}{\langle h(t_0)\rangle}
,
\label{eq:lifetime}
\end{equation}
where $h(t)$ is a dynamical variable, defined as 1 if a specific pair of atoms is bonded at time $t$, and 0 otherwise. To identify a bond, we used the bonding cutoff defined in Ref. \cite{omar2024} from inspection of the pair correlation functions, namely 3.2 {\AA } for all pairs except for Sb–Te for which we use 3.4 {\AA }.

The intermittent bond lifetime $\tau_b$ is defined as the time it takes for the autocorrelation function $C(t)$ to decay to $1/e$ of its initial value at $t$=0.  However, due to aging, the lifetime actually depends on time in our simulations. Therefore,  we analyzed $C(t)$ in a time much shorter than $\tau_b$ at which it can be approximated by 

\begin{equation}
C(t) \simeq e^{-t/\tau_b} \simeq 1 -t/\tau_b. 
\label{eq:lifetimeshort}
\end{equation}
The autocorrelation function $C(t)$ resolved for the different pairs  is shown in Fig. \ref{C_t} on a time scale of 7 ns at the beginning and end of the simulation, where $C(t)$ has been averaged over the initial times (\ref{eq:lifetime}) in a 3 ns interval. At the beginning of the simulation, there is a rapid change in
$C(t)$ probably related to a fast relaxation. The lifetime was then extracted using Eq. \ref{eq:lifetimeshort}, which yields the bond lifetimes
given in Table \ref{table:lifetime}.

As already found in liquid and supercooled liquid GST (700-1100 K)\cite{lee2025wrongbonds} wrong  Ge-Ge, Ge-Sb, and Te-Te bonds have shorter lifetimes than Ge-Te and Sb-Te bonds, with Te-Te being the shortest lived bond. The lifetime of Sb-Sb bond is instead  very close  to the Ge-Te one as also found at higher temperatures in Ref. \cite{lee2025wrongbonds}. Lifetimes increase by a factor three or more from the beginning to the end of the simulation, indicating that structural relaxations due to aging lead to a more stable and less mobile bonding network.

The lifetimes in Table \ref{table:lifetime} indicate that most of the bond breaks in the timescale of our simulations, which is also consistent with the large displacement of a subset of atoms (see  Fig. \ref{DP}a and the discussion therein).
However,  the distribution of the different types of bonds changes very little from the beginning to the end of the simulations, as we now discuss in the next section.

\subsection{Structural changes upon aging}\label{structure}
\begin{figure}
    \centering
    \includegraphics[width=0.38\linewidth,keepaspectratio]{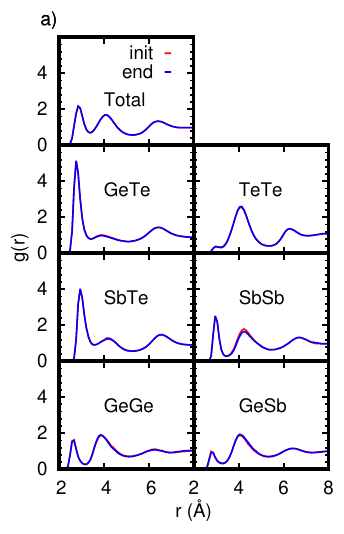}   \includegraphics[width=0.29\linewidth,keepaspectratio]{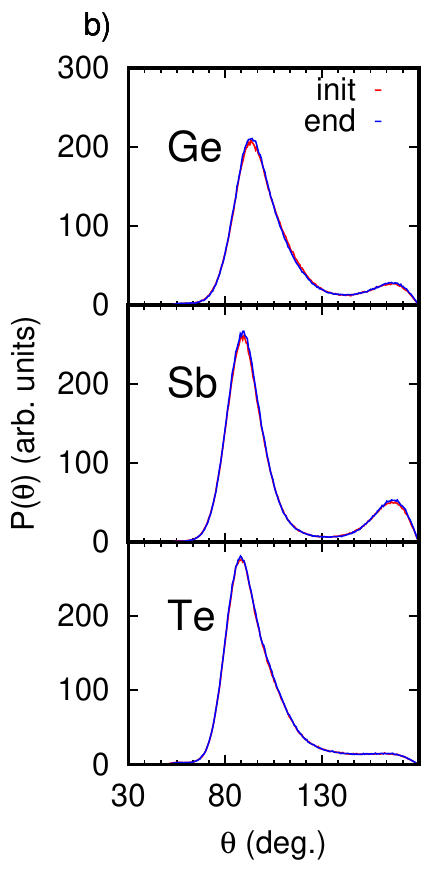}
    \caption{a) Partial pair correlation functions and b) bond angle distribution at beginning  and at the end of the simulation of the large model at 450 K, averaged over 100 ps. } 
    \label{fig:gofrA}
\end{figure}

\begin{figure} [h!]
    \centering
    \includegraphics[width=0.55\linewidth,keepaspectratio]{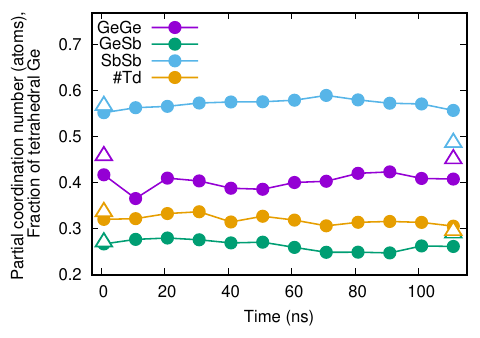}
    \caption{Evolution in time of the partial coordination number corresponding to wrong bonds and fraction of Ge atoms in tetrahedral coordination in the simulation of the large model at 450 K. Each point corresponds to an average over 100 ps. The triangles at the beginning and end of simulation refer to the subset of MM atoms defined by the DP (see Fig. \ref{DP}).} 
    \label{fig:coord}
\end{figure}

The partial pair correlation functions and the bond angle distribution function at the beginning and end of the simulation  are nearly indistinguishable, as shown in Fig. \ref{fig:gofrA}. 
The partial coordination numbers related to the fraction of wrong bonds and the fraction of tetrahedral Ge atoms change also very little in the time span of 100 ns, as shown  in Fig. \ref{fig:coord}. 
The partial coordination numbers are obtained by integrating the partial pair correlation functions up to the 
bonding cutoff of 3.2 {\AA } for all pairs except for Sb–Te for which we use 3.4 {\AA }  as we did in Ref. \cite{omar2024}.
Even if we restrict the calculation of the partial coordination numbers to the MM atoms (see Fig. \ref{DP}a) no major modulation of the number of wrong bonds with time is found (triangles in Fig. \ref{fig:coord}), but for a small decrease of the fraction of Sb-Sb bonds.

Apparently, the structural transformations that reduce the number of wrong bonds and tetrahedra, and then the in-gap states according to the analysis of Ref. \cite{OmarActa}, should occur on a time scale much longer than the 100 ns of our simulations.

\begin{figure}
    \centering  \includegraphics[width=0.38\linewidth,keepaspectratio]{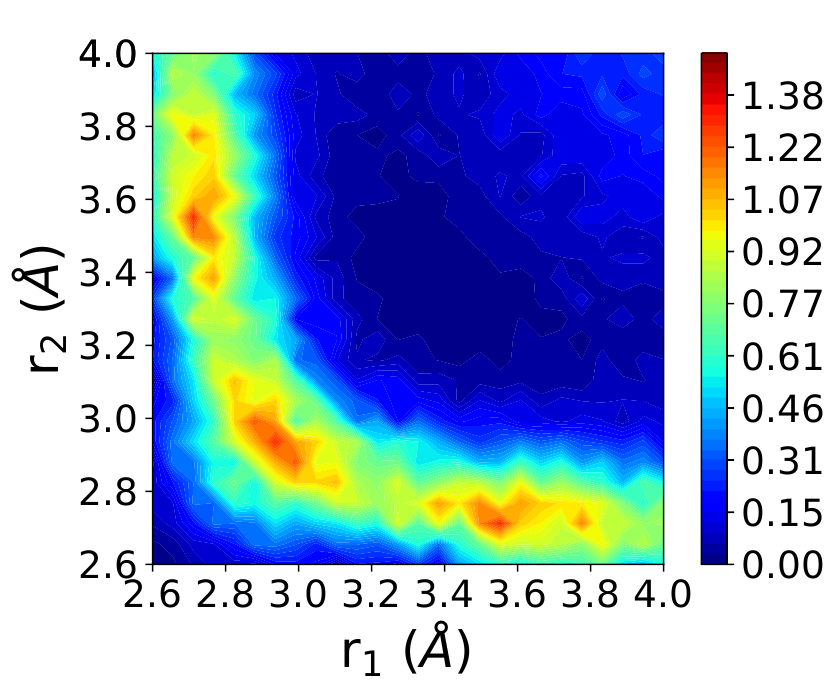}
\includegraphics[width=0.4\linewidth,keepaspectratio]{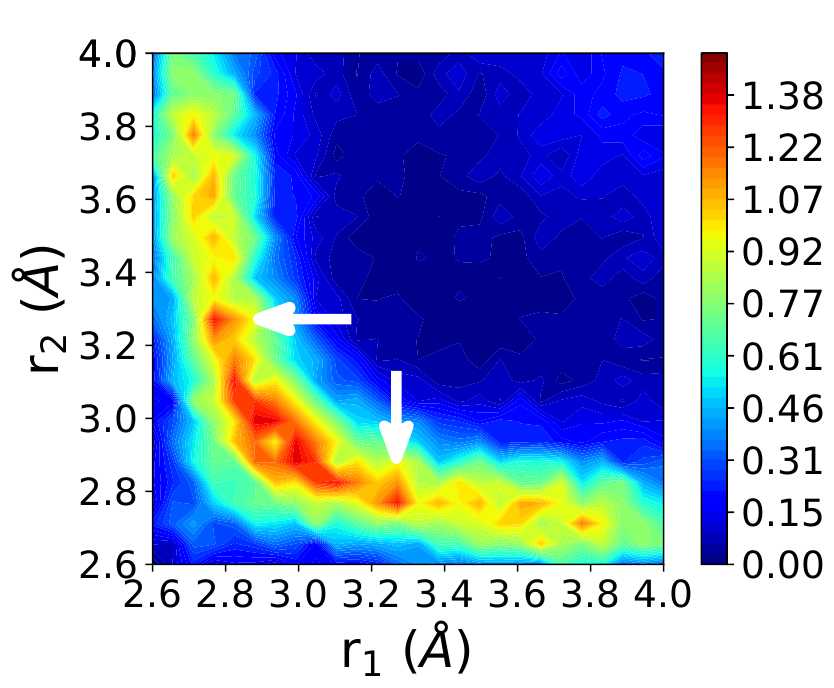}
    \caption{ALTBC function for Ge atoms at the beginning (left) and at the end (right) of the simulation of the large model at 450 K. The analysis is restricted to the MM atoms (see Fig. \ref{DP}a and the discussion therein). The ALTBC is averaged over 50 ps (one frame per ps).
    The arrows in the right panel indicate the secondary peaks corresponding to long-short alternation of axial bonds.} 
    \label{fig:ALTBC}
\end{figure}

On the other hand, aging leads to an enhancement of the Peierls distortion that we quantified by computing the 
 angular-limited three-body correlation function (ALTBC) \cite{Bichara} which highlights the presence of a short and a long axial bond in defective octahedral configurations. Namely, ALTBC measures the probability of finding
 a bond with length $r_1$ mostly aligned (within a threshold angle of 25$^o$)  with a second bond of length $r_2$ formed with the same central atom.
 The ALTBC function in the ($r_1$, $r_2$) plane is shown in Fig. \ref{fig:ALTBC} for central Ge atom at the beginning and at the end of simulation A, by considering only the subset of MM atoms.
At the end of the simulation there is a clear enhancement of the secondary peaks (indicated by arrows in Fig. \ref{fig:ALTBC}) corresponding to long-short alternation of axial bonds. This effect is overall small and not visible in the global ALTBC involving all atoms, but emerges once the analysis is restricted to the atoms that move the most. This means that the enhancement of the  Peierls distortion  does not require just a small local rearrangement, but that it occurs after more extended structural transformations.

\subsection{Changes in electronic properties upon aging}\label{electronic}
\begin{figure}
    \centering
    \includegraphics[width=0.45\linewidth,keepaspectratio]{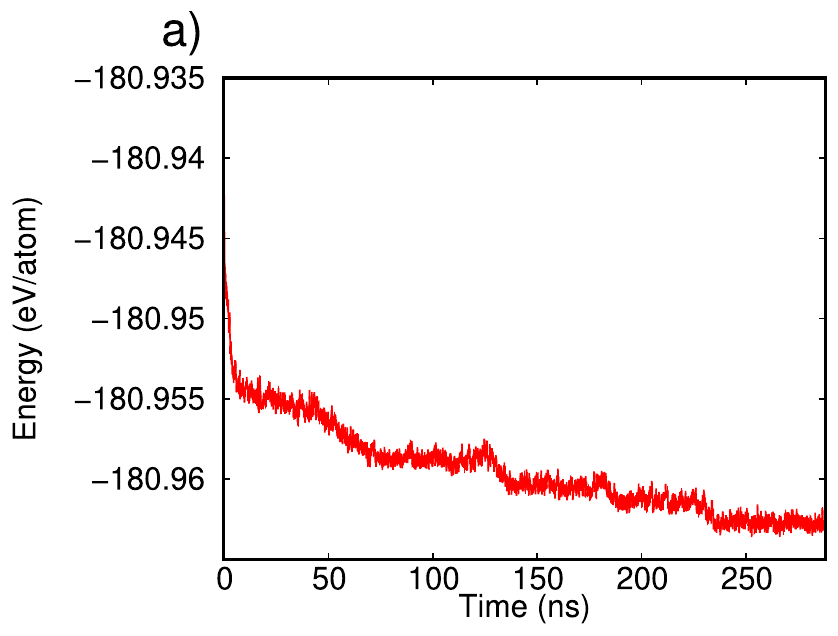}
    \includegraphics[width=0.4\linewidth,keepaspectratio]{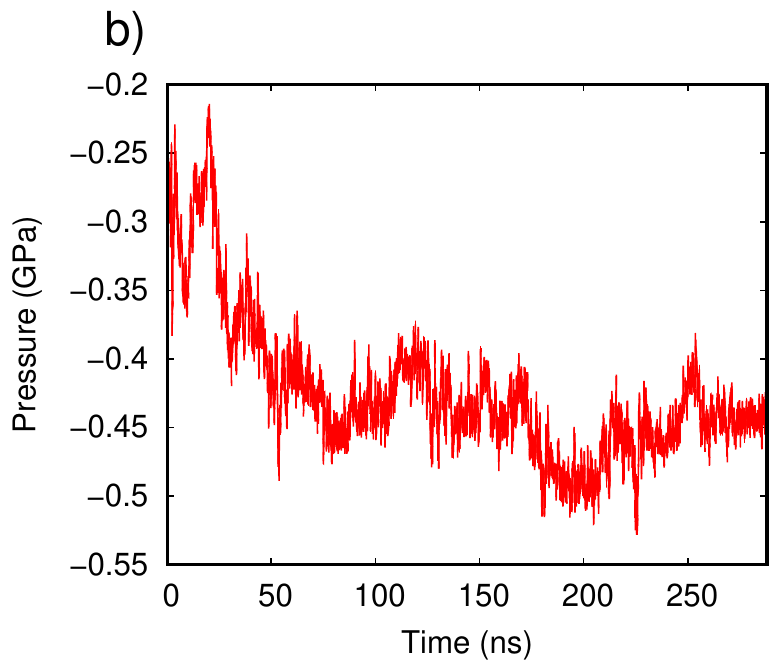}
    \caption{a) Energy per atom and b) pressure over time in the simulation of the small model at 400 K.}
    \label{EnergySmall}
\end{figure}

To analyze the effect of aging on the electronic defects, we considered a smaller 999-atom model on which electronic structure calculation can be performed within the DFT framework described in Sec. 2. 
 We performed a MLIP-MD NVT simulation 288 ns long at 400 K whose evolution in time of the energy and pressure is shown in Fig. \ref{EnergySmall}. We observe a rapid drop in energy in the first few ns and then a slower decrease over time. The same rapid drop was observed  in the long equilibration  run at 400 K of the large model, not shown in Fig. \ref{energyA} that refers to the subsequent aging  run at 450 K.
 Consistently with what we did for the large model, we analyzed aging after 20 ns of equilibration at 400 K, i.e. after the rapid drop in energy. 
 We consider this point the beginning of the aging simulation, as shorter times would correspond to relaxations of the glass during the reset process of the memory as discussed in Sec. 2.

\begin{figure}
    \centering
    \includegraphics[width=0.6\linewidth]{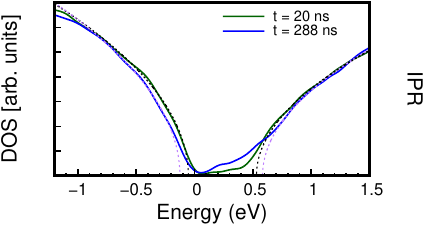}
    \caption{Electronic density of states (DOS) at the HSE level averaged over 10 snapshots (see text) of two DFT-PBE simulations starting either from the beginning or the end of aging simulation of the small model at 400 K. The DOS is shown close to the band gap and is fitted at the two edges with the function $\sqrt{|\varepsilon - \varepsilon_{v/c}|}$, where $\varepsilon_{v/c}$ is the edge of the valence/conduction band.
    The DOS of the different snapshots of the MD trajectories at 300 K (see text) are aligned to the lowest energy state due to 5s orbitals of Te which keep an atomic like character with small dispersion in energy. 
    The average DOS at the initial and final configurations are then aligned to the highest occupied KS state (zero of energy).
    Each KS energy is broadened with a Gaussian function with a 54 meV variance.} 
    \label{DOS-C}
\end{figure}

Then, we extracted a snapshot at the beginning and at the end of this aging simulation. From these two configurations, we first optimized the models at 0 K with the MLIP to generate an amorphous model. Then we equilibrated the amorphous model at 300 K in a DFT-PBE MD simulation  20 ps long, each. 
 From each of the latter two MD runs, we extracted 10 snapshots 
 whose geometry was then optimized at 0 K  with the hybrid HSE \cite{hse06} functional to better reproduce the band gap. The snapshots were extracted after 2 ps of the DFT-PBE NVT run when the total energy stabilized.  
 The resulting DOS at the HSE level at the beginning and at the end of the aging simulation,
 averaged over the 10 snapshots,  is shown in Fig. \ref{DOS-C}.  We performed DFT-PBE MD simulations instead of starting directly from DFT-HSE MD simulations because of the higher computational cost of the latter.

 \begin{figure}
    \centering
    \includegraphics[width=0.45\linewidth]{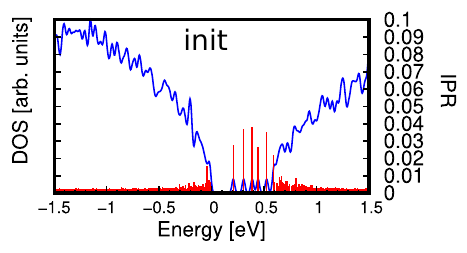}
     \includegraphics[width=0.45\linewidth]{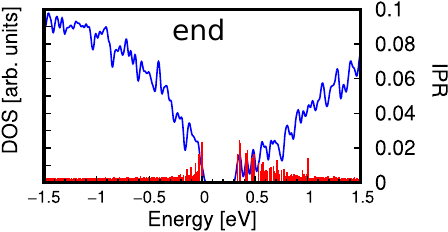}
    \caption{Electronic density of states (DOS) superimposed to the Inverse Participation Ratio (IPR) of a single snapshot at the beginning (initial) and at the end of the simulation of the small model at 400 K after geometry optimization with the HSE functional. Each KS energy is broadened with a Gaussian function with a 13.6 meV variance.   
    } 
    \label{DOS-IPR-C}
\end{figure}

The DOS has a finite value within the band gap due to the presence of localized states,  on which we will comment later on. 
The DOS close to the gap was fitted by a function $\sqrt{|\varepsilon - \varepsilon_{v/c}|}$, where $\varepsilon_{v/c}$ is the edge of the valence/conduction band, as shown in Fig. \ref{DOS-C}. After aging, there is a clear widening of the band gap ($\varepsilon_c - \varepsilon_{v}$) that results from both  a blue-shift of $\varepsilon_c$ ad a red-shift of $\varepsilon_v$. Note that the DOS are aligned to the lowest energy state due to 5s orbitals of Te which keep an atomic like character with small dispersion in energy. 
The band gap increases from 0.582 eV to 0.714 eV after 288 ns aging. According to the analysis of Sec. \ref{structure}, we can link the widening of the band gap with the enhancement of the Peierls distortion, as also suggested in \cite{Raty2015} from the analysis of different models of a-GeTe. However, in the small 999-atom cell the MM atoms defined as given above for the large cell, are relatively few, and the statistics is not sufficient to get a clear signal out of the noise in the ALTBC. Still, the effect of the enhanced Peierls distortion that we expect to take place in the small cell as well, on the basis of the results on the larger cell, is visible in the electronic DOS.

As mentioned above, the DOS is finite inside the gap due to the presence of localized states.
The DOS of  two individual snapshots at the beginning and at the end of the aging simulation are compared in Fig. \ref{DOS-IPR-C}, superimposed to the IPR (see Sec. II) of individual KS states.
The DOS close to the band gap and the IPR of the 10 individual snapshots at the beginning and end of the aging simulation are shown in Figs. S1-S2 in the Supplementary Material.
In a recent work \cite{Drabold}, it is shown that  the position in energy and the degree of localization of in-gap states in amorphous GST are subject to large fluctuations at 300 K on the ps time scale in  DFT-HSE MD simulations.
Similarly, large fluctuation in the in-gap states are found here in the DOS and IPR of the models optimized at the HSE level at 0 K starting from different snapshots of the DFT-PBE simulations at 300 K (Figs. S1-S2 in the SM). However, by comparing the average DOS (over 10 snapshots) at the beginning and at the end of the simulation in Fig. \ref{DOS-C}, it is clearly seen that the widening of the 
band gap is accompanied by an increase of the Urbach tail of localized states at the edge of the conduction band.

\section{Conclusions}
In summary, we have simulated the aging of amorphous Ge$_2$Sb$_2$Te$_5$ by brute force MD simulations with a machine learning interatomic potential previously devised \cite{omar2024}. We analyzed aging at 450 K, slightly below the experimental T$_g$, in a 100 ns simulation of a 7992-atom model after equilibration for 24 ns at a slightly lower temperature of 400 K. First, we observed  a small reduction of stress in the time scale of a few tens of ns in fixed volume simulations, aimed at mimicking the conditions of the memory device. Instead, the energy  continuously decreased with a slower pace at longer times. Upon aging, we observed a reduction in atomic mobility that is highly heterogeneous with a fraction of atoms that undergo displacements much longer than the average, as quantified by the distribution of the dynamical propensity and by the non-Gaussianity parameter $\alpha^2$. A large fraction of bonds break and reform on the simulation time scale. The bond lifetime computed from the decay of the bond-bond correlation function increases upon aging from 12–27 ns at the beginning to 39–90 ns at the end of the simulation. Wrong bonds (homopolar and GeSb) and especially Te-Te bonds have shorter lifetimes as also observed in the liquid \cite{lee2025wrongbonds}.

In spite of a substantial atomic motion, the fraction of wrong bonds, which are supposed to be absent in the ideal aged glass, does not change appreciably in 100 ns. Nor the fraction of tetrahedral Ge sites does.
Apparently, wrong bonds, which are also expected to control part of the in-gap states responsible for the subthreshold conduction, could be removed only on a larger time scale. In contrast, in 100 ns we observed an enhancement of the Peierls distortion that well correlates with the widening of the electronic band gap observed upon aging in a smaller 999-atom cell for which DFT-HSE calculations have been performed. The widening of the band gap was accompanied by an increase in localized states in the Urbach tail of the conduction band edge that could eventually be removed only at longer times. Therefore, the simulations suggest a sequence of transformations on different time scales: first a rapid relaxation of the stress on a few tens of ns, then an enhancement of the Peierls distortion responsible for the widening of the band gap and an overall slowdown of the bond breaking dynamics on the time scale of 100 ns.
A further transformation into an ideal glass with the removal of wrong bonds and of the accompanying in-gap states \cite{OmarActa} is expected to occur on a longer time scale that requires accelerated methods to be tackled by MD simulations  such as the metadynamics used in previous work \cite{OmarActa}.

\section*{Funding}
The project has received funding from European Union Next-Generation-EU  through the Italian Ministry of University and Research under PNRR M4C2I1.4 ICSC 371 Centro Nazionale di Ricerca in High Performance Computing, Big Data and Quantum 372 Computing (Grant No. CN00000013). 
We acknowledge the CINECA award under the ISCRA initiative, for the availability of high-performance computing resources and support.

\section*{Data availability}
LAMMPS, DeePMD and CP2k are free and open source codes available at https://lammps.sandia.gov, http://www.deepmd.org and  https://www.cp2k.org, respectively. The MLIP developed in Ref. \cite{omar2024} is available in the Materials Cloud repository at https://doi.org/10.24435/materialscloud:a8-45.
Trajectories of the simulations can be made available upon request.

\bibliographystyle{iopart-num}
\bibliography{ref}

\onecolumngrid
\newpage

\vspace*{\fill}
\begin{center}
\noindent\Huge{Supplementary Material} 
\end{center}
\vspace*{\fill}

 \setcounter{figure}{0}

\begin{figure}
    \centering
             \renewcommand\figurename{Figure~S$\!\!$}
\includegraphics[width=0.4\linewidth]{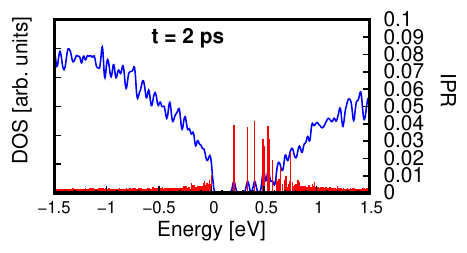}
\includegraphics[width=0.4\linewidth]{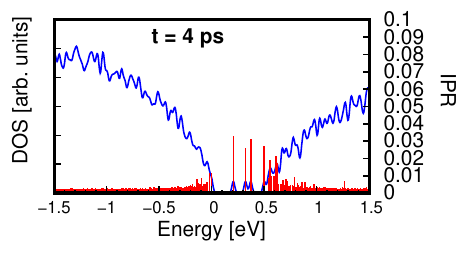}
\includegraphics[width=0.4\linewidth]{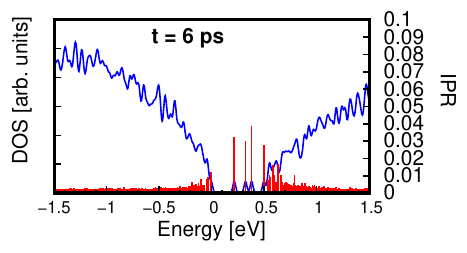}
\includegraphics[width=0.4\linewidth]{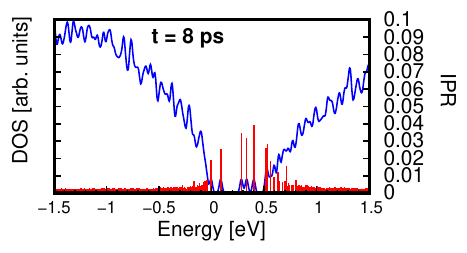}
\includegraphics[width=0.4\linewidth]{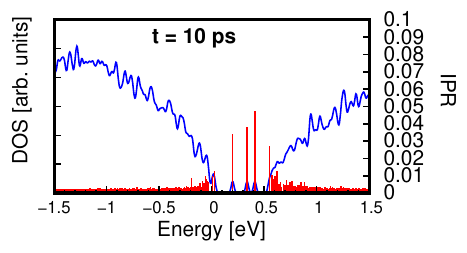}
\includegraphics[width=0.4\linewidth]{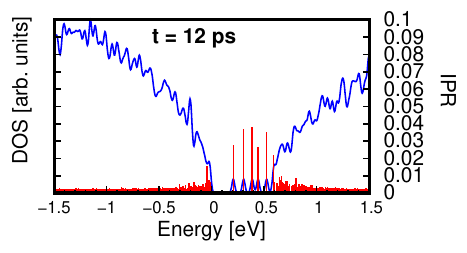}
\includegraphics[width=0.4\linewidth]{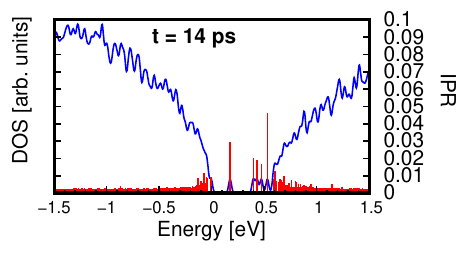}
\includegraphics[width=0.4\linewidth]{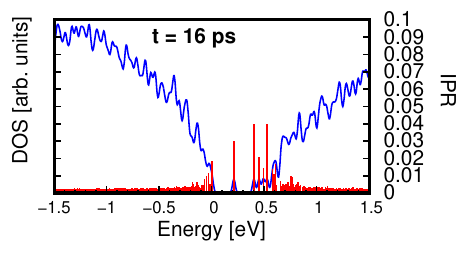}
\includegraphics[width=0.4\linewidth]{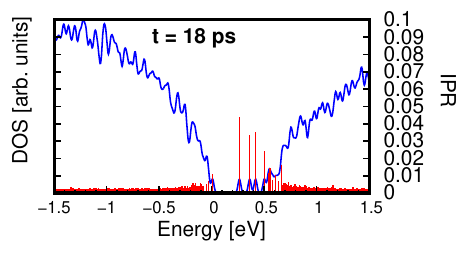}
\includegraphics[width=0.4\linewidth]{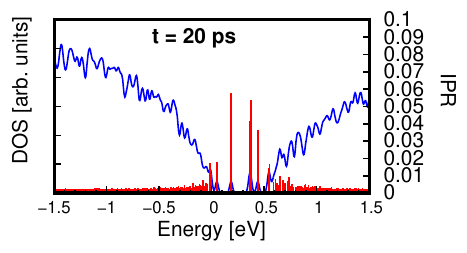}
    \caption{Electronic density of states (DOS) superimposed to the Inverse Participation Ratio (IPR) of 10 snapshots at the beginning of simulation of the small model (999-atom at 400 K). The HSE functional was used. The time given in the inset corresponds to those of the short DFT-PBE from which the snapshots are extracted and optimized at the HSE level. Each KS energy is broadened with a Gaussian function with a 13.6 meV variance.} 
    \label{DOS-IPR-C-INI}
\end{figure}


\begin{figure}
    \centering
             \renewcommand\figurename{Figure~S$\!\!$}
    \includegraphics[width=0.4\linewidth]{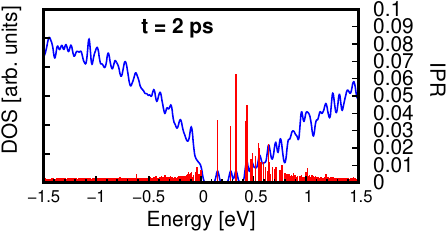}
    \includegraphics[width=0.4\linewidth]{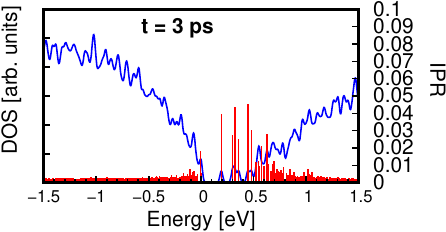}
\includegraphics[width=0.4\linewidth]{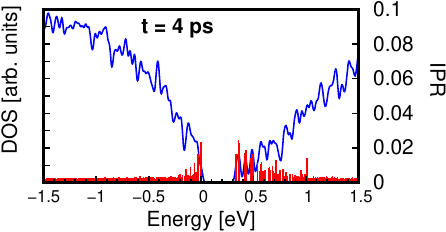}
\includegraphics[width=0.4\linewidth]{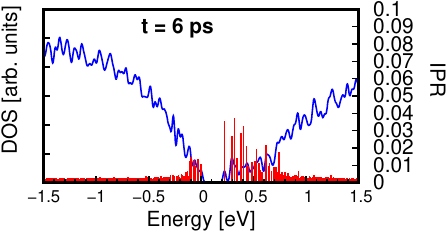}
\includegraphics[width=0.4\linewidth]{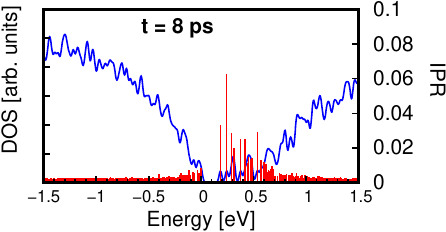}
\includegraphics[width=0.4\linewidth]{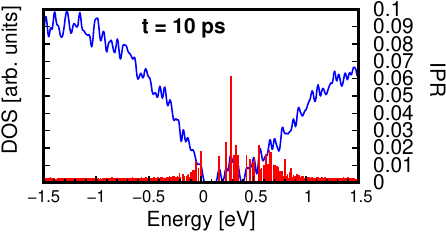}
\includegraphics[width=0.4\linewidth]{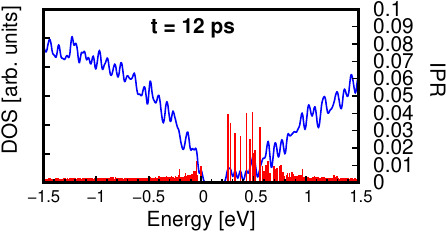}
\includegraphics[width=0.4\linewidth]{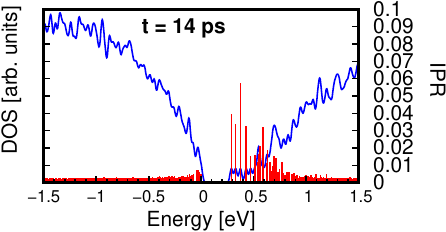}
\includegraphics[width=0.4\linewidth]{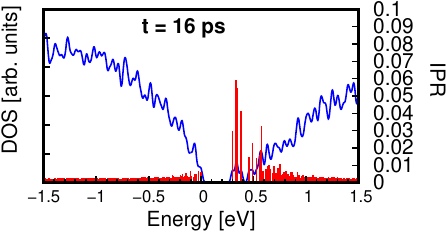}
\includegraphics[width=0.4\linewidth]{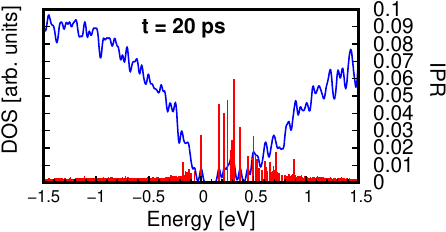}
  \caption{Electronic density of states (DOS) superimposed to the Inverse Participation Ratio (IPR) of 10  snapshots at the end of the simulation of the small model (999-atom at 400 K). The HSE functional was used. The time given in the inset corresponds to those of the short DFT-PBE from which the snapshots are extracted and optimized at the HSE level. Each KS energy is broadened with a Gaussian function with a 13.6 meV variance.} 
    \label{DOS-IPR-C-END}
\end{figure}

\end{document}